\documentclass[10pt,twocolumn,twoside]{IEEEtran}

\usepackage[lined,boxed,commentsnumbered,ruled]{algorithm2e}
\usepackage{amsfonts}
\usepackage{mathrsfs}
\usepackage{amssymb,amsmath}

\usepackage{amsmath,amssymb,epsfig,graphics,subfigure}
\usepackage{theorem}
\usepackage{array,color}
\usepackage[compress]{cite}

\begin{document}

% ==================================================
\title{Distributed Resource Allocation Algorithm Design for Multi-Cell Networks Based on Advanced Decomposition Theory}

\author { Zesong Fei$^{*}$ , Shuo Li, Chengwen Xing, Yiqing Zhou and Jingming Kuang \\
\thanks{Z. Fei, S. Li, J. Kuang and C. Xing are with RCDCT, Modern Communication Lab,
Dept. of E. E., Beijing Institute of Technology, Beijing, China (email: \{feizesong, surelee, jmkuang\}@bit.edu.cn; chengwenxing@ieee.org)

Y. Zhou is with Wireless Communication Research Center,
Institute of Computing Technology, Chinese Academy of Sciences, Beijing, China (email: zhouyiqing@ict.ac.cn)}}

%\markboth{Journal of \LaTeX\ Class Files,~Vol.~6, No.~1, January~2007}%
%{Shell \MakeLowercase{\textit{et al.}}: Bare Demo of IEEEtran.cls for Journals}

\maketitle

% ==================================================
\begin{abstract}
In this letter, we investigate the resource allocation for downlink multi-cell coordinated OFDMA wireless networks, in which power allocation and subcarrier scheduling are jointly optimized. Aiming at maximizing the weighted sum of the minimal user rates (WSMR) of coordinated cells under individual power constraints at each base station, an effective distributed resource allocation algorithm using a modified decomposition method is proposed, which is suitable by practical implementation due to its low complexity and fast convergence speed. Simulation results demonstrate that the proposed decentralized algorithm provides substantial throughput gains with lower computational cost compared to existing schemes.
\end{abstract}


% ==================================================
\section{Introduction}
%Orthogonal frequency-division multiple access (OFDMA) has been taken as a promising radio access technique due to its strong ability to support broadband wireless communications. To improve spectrum efficiency, universal frequency reuse is always assumed in OFDMA-based wireless cellular systems, which causes serious co-channel interference (CCI) in a multicell system. Joint resource allocation (RA) among coordinated BSs has emerged as a promising way for co-channel interference (CCI) mitigation and system optimization\cite{Wang2009}.

OFDMA has been widely accepted as a promising multiple access technique for future mobile communication systems. Recently, OFDMA-based multi-cell coordination becomes a hot research topic, which could provide superior performance over the traditional single-cell processing network through joint signal processing among the involved based stations \cite{Wang2009}. Moreover, the performance of the multi-cell coordination network can be further improved by employing multi-cell resource allocation.

However, the traditional centralized multi-cell resource allocation approach is too complicated to be practical because of the large number of parameters and constraints \cite{KTH}. Therefore, various distributed algorithms have been proposed to reduce the computation complexity and increase the scalability, such as the distributed algorithm for the resource allocation using game theory in \cite{game} and the one based on the Lagrange duality method in \cite{Zhangrui}.

In this paper, a novel distributed resource allocation algorithm is proposed for multi-cell OFDMA networks. Different from existing works in which weighted sum of user rates (WSR) are usually used, the weighted sum of the minimal user rates (WSMR) of coordinated cells is taken as the performance criterion, which should be maximized subject to individual power and subcarrier allocation constraint at each BS. An iterative algorithm is proposed to solve the problem, which optimizes the subcarrier scheduling and the power allocation alternatively. For the power allocation subproblem, motivated by a natural decomposition of the optimality conditions of the original problem, we propose a novel distributed algorithm. Unlike traditional Lagrangian based algorithms, the proposed technique does not need to solve subproblems, which results in computational savings and fast convergence. Furthermore, in the proposed procedure the central agent only distributes information and checks the convergence condition without the need to update information as \cite{Zhangrui}, which makes the scheme simpler.

%The rest of this paper is organized as follows. First, we present the system model and the problem formulation for the considered system. In Section~\ref{sect:system}, we present the proposed distributed resource allocation algorithm. We describe the general framework of the proposed algorithm firstly, and then investigate the power allocation subproblem and subcarrier allocation subproblem in turn. In Section~\ref{sec:simulation}, we illustrate the effectiveness of the proposed algorithm through numerical experiments. Finally, Section VI concludes the paper.

\section{System Model And Problem Formulation}
\label{sect:system}

\subsection{System Model}

In this paper, we investigate the distributed resource allocation in the downlink of a cellular OFDMA network consisting of $M$ cooperative cells. In the $m^{\rm{th}}$ cell, BS $m$ serves $K_{m}$ users and the active user set is denoted as $U_m$. The channel power coefficient of subcarrier $n$ from BS $l$ to user $u$ in the $m^{\rm{th}}$ cell is denoted by $g^{l}_{u,m,n}$ . If $l \neq m$, $g^{l}_{u,m,n}$ is related to the interfering channel power from BS $l$ , and $g^{l}_{u,m,n}$ denotes the desired channel power coefficient from BS $m$ if $l=m$. In addition, $\sigma^{2}_{u,m,n}$ is the power of the additive white Gaussian noise at user $u$ in cell $m$ on subcarrier $n$. The number of OFDMA subcarriers is $N$ and each subcarrier is allocated to only one user exclusively in each cell. It is also assumed that all channel state information (CSI) is perfectly known at each BS.

The maximum transmit power of BS $m$ is denoted as $P_{m,max}$, and $P_{m,n}$ represents the power allocated to subcarrier $n$ by BS $m$. For convenience, $\{P_{m,n}\}^{M}_{m=1}$ is stacked into a $M \times 1$ vector $\mathbf{p}_{n}=[ P_{1,n},\ldots,P_{M,n}]^{T}$, and then $\{\mathbf{p}_{n}\}^{N}_{n=1}$ is stacked into a $M \times N$ matrix $\mathbf{P}$ referring to power allocation in following sections. As for the subcarrier allocation notations, we define a binary variable $A_{u,m,n}$, which indicates that subcarrier $n$ is allocated to user $u$ in cell $m$ if $A_{u,m,n}=1$. Stack $\{A_{u,m,n}\}^{N}_{n=1}$ into a $N \times 1$ vector $\mathbf{A}_{u,m}=[A_{u,m,1},\ldots,A_{u,m,N}]^{T}$, and merge $\mathbf{A}_{u,m}$ of all users in all cells to a $N \times K_{m} \times M$ subcarrier allocation matrix $\mathbf{A}$ which indicates how subcarriers are assigned among all users.

Then the information rate of user $u$ in cell $m$ can be expressed as a function of $\mathbf{A}_{u,m}$ and $\mathbf{P}$
\begin{equation} \label{user rate with A}
\begin{split}
R_{u,m}(\mathbf{A}_{u,m},\mathbf{P}) &= {\sum}_{n=1}^N A_{u,m,n} \cdot R_{u,m,n}(\mathbf{p}_n)\\
%                                     &= \sum_{n:A_{u,m,n=1}}R_{u,m,n}(\mathbf{p}_n)
\end{split}
\end{equation}
where $R_{u,m,n}$ is the achievable rate of user $u$ on subcarrier $n$ in cell $m$, and can be calculated as
\begin{equation} \label{user rate without A}
R_{u,m,n}(\mathbf{p}_n) = \textrm{ln} \left(1+ \frac{P_{m,n}g^{m}_{u,m,n}}{(\sigma^{2}_{u,m,n}+\sum_{l=1,l\neq m}^M P_{l,n}g^{l}_{u,m,n})\Gamma}\right)
\end{equation}
in the unit of nats per OFDM symbol,
where $\gamma_{u,m,n}(\mathbf{p}_n)$ is the signal to interference plus noise ratio (SINR), and $\Gamma$ represents the signal to
noise ratio gap between the adopted modulation and coding scheme and the one achieving capacity.
%\begin{equation} \label{SNIR}
%\gamma_{u,m,n}(\mathbf{p}_n) = \frac{P_{m,n}g^{m}_{u,m,n}}{\sigma^{2}_{u,m,n}+\sum_{l=1,l\neq m}^M P_{l,n}g^{l}_{u,m,n}}.
%\end{equation}

\subsection{Problem Formulation}

For resource allocation, sum capacity is the most widely used performance criterion. However, since OFDMA multi-cell networks with multiple users in each cell, maximizing the sum capacity of a multi-cell network may result in a serious performance imbalance. It is because that the users and cells with better channel conditions will be allocated with much more resources and those experiencing worse channels may be sacrificed. In order to overcome this kind of problems fairness among the users and cells must be taken into account.

Therefore, an objective function called weighted sum minimal user rate (WSMR) is used. As shown in \cite{KIM}, maximizing the minimal user rate in a single cell could provide the maximum fairness among the users in this cell. Therefore, in cooperative multi-cell systems, by introducing different priorities to the involved cells, WSMR can be employed to provide fairness to multiple users as well as cells.
The objective function in this problem is given by
\begin{equation}
\label{object function}
 f(\mathbf{A,P}) = {\sum}_{m=1}^M \omega_m \cdot \min_{u \in U_m} R_{u,m}(\mathbf{A}_{u,m},\mathbf{P})
\end{equation}
where $\omega_m \geq 0$ represents the weight assigned to cell $m$'s minimal user rate. In particular, increasing $\omega_m$ leads to a higher resource allocation priority assigned to the users in cell $m$. Based on (\ref{object function}), the resource allocation problem is formulated as
\begin{align}
\label{original optimization problem}
 \max_{\mathbf{A,P}}&\quad f(\mathbf{A,P}) = {\sum}_{m=1}^M \omega_m \cdot \min_{u \in U_m} R_{u,m}(\mathbf{A}_{u,m},\mathbf{P}) \nonumber \\
 \textrm{s.t}&  \quad {\sum}_{n=1}^N P_{m,n} \leq P_{m,max} \ \ \  \forall{m}, \nonumber  \\
             & \quad {\sum}_{u \in U_m}A_{u,m,n} \leq 1  \ \ \ \forall{m,n},\nonumber  \\
             & \quad A_{u,m,n} \in \{0,1\} \quad\quad\forall{u, m, n}.
\end{align}

 Note that the minimization operation in (\ref{original optimization problem}) prohibits the objective function from being differentiable. Thus, an auxiliary optimization variable $R_m$ is introduced, and the optimization problem (\ref{original optimization problem}) is reformulated as follows
\begin{align} \label{equivalent optimization problem}
 \max_{\mathbf{A,P},R_m}&\
\ \  {\sum}_{m=1}^M \omega_m \cdot R_m \nonumber \\
 \textrm{s.t}&\quad {\sum}_{n=1}^N A_{u,m,n} \cdot R_{u,m,n}(\mathbf{p}_n) \geq  R_m \ \ \forall{m,u \in U_m}\nonumber \\
             &\quad {\sum}_{n=1}^N P_{m,n} \leq P_{m,max} \nonumber \\
             & \quad {\sum}_{u \in U_m}A_{u,m,n} \leq 1 \ \  \forall{m,n}\nonumber \\
             &\quad A_{u,m,n} \in \{0,1\} \ \  \forall{m,n,u \in U_m}.
\end{align}

Obviously, the optimization problem (\ref{equivalent optimization problem}) is a constrained nonlinear optimization program with both integer and continuous variables. Furthermore, the problem is not convex and thus generally speaking it is difficult to directly solve the problem. Following a similar logic as those in \cite{Wong1999,Rhee2000},  an effective algorithm is proposed in the following, which optimizes subcarrier scheduling and power allocation alternatively.
\section{The Proposed Algorithm}
\label{sect:Algorithm }

\subsection{Proposed Decomposition Power Allocation Algorithm}
Without loss of generality, the power allocation optimization is carried out first at each iteration. The subcarrier allocation variables \textbf{A} can be directly removed from (\ref{equivalent optimization problem}) as they are assumed to be already computed.  For notational simplicity, all superscript of variables are omitted in following description. Thus, the power allocation optimization subproblem can be expressed as
\begin{align}
\label{power allocation optimization problem}
 \max_{\mathbf{P},R_m} \ & \ \  \ \  {\sum}_{m=1}^M \omega_m \cdot R_m \nonumber \\
 \textrm{s.t} & \ \ \ \  {\sum}_{{\tiny{n:A_{:,:,n=1}}}}{R_{u,m,n}(\mathbf{p}_n)} \ge   R_m  \nonumber \\
 & \ \ \ \  {\sum}_{n=1}^N P_{m,n} \leq P_{m,max}.
\end{align}
Based on the definition of $R_{u,m,n}(\mathbf{p}_n)$ in (\ref{user rate without A}), it is obvious that (\ref{power allocation optimization problem}) is a nonconvex optimization problem. For multi-cell OFDMA networks, the optimization problem (\ref{power allocation optimization problem}) usually has high dimensions. Although centralized optimization methods can be used to solve the problem, they are impractical due to the high complexity. Therefore, instead of centralized optimization algorithms, we focus on distributed power allocation schemes using decomposition techniques.

Various decomposition algorithms have been developed. The frequently used decomposition algorithm in wireless communications is Lagrangian relaxation procedure \cite{Bazaraa1993,Kelley1960}, and its variant relaxation techniques based on augmented Lagrangian functions \cite{Carpentier1996,Cohen1978}.
However, Lagrangian procedures may present drawbacks in some cases, such as difficulties to converge to an optimal solution for the global system (in the absence of convexity assumptions), uncontrollable convergence rates that depend on the correct choice of the values for several parameters which are difficult to update, and the requirement of the intervention of a central agent to update this complicated information.

To overcome these drawbacks, a novel decomposition algorithm which is based on the decomposition of the optimality conditions of the global problem (\ref{power allocation optimization problem}) is proposed in this paper \cite{Bazaraa1993}. The proposed decomposition algorithm improves both the computational efficiency and implementation simplification compared to previously mentioned algorithms. In the following, the decomposition methodology is presented in detail.

%\noindent \underline{\textbf{Proposed decomposition methodology:}}

First of all, the optimization problem (\ref{power allocation optimization problem}) can be written in a compact form as follows for convenience
\begin{align} \label{compact form of power subproblem}
 \max_{\boldsymbol{x}_m}&\quad {\sum}_{m=1}^M f_m(\boldsymbol{x}_m) \nonumber \\
 \textrm{s.t}&\quad \boldsymbol{h}(\boldsymbol{x}_1,\dots,\boldsymbol{x}_M) \leq \boldsymbol{0} \quad \boldsymbol{g}_m(\boldsymbol{x}_m) \leq \boldsymbol{0}
\end{align}
where $\boldsymbol{x}_m$ is a vector for cell $m$, that contains power allocation vector $\mathbf{p}_m$ and the minimum user rates $R_m$ within cell $m$.

The first constraint in (\ref{compact form of power subproblem}) is known as complicating constraint, which represents the rate constraint in (6). These equations contain variables and parameters from different cells and prevent each system from operating independently to each other. On the contrary, the second constraint is the power constraint for each BS and thus only related to one single cell. Therefore, in order to decompose the global optimization problem (\ref{compact form of power subproblem}) into local subproblems for each cell, those equations of rate constraints are removed from (\ref{compact form of power subproblem}). Thus, the problem is equivalent to
\begin{align} \label{compact form of power subproblem-decomposed}
 \max_{\boldsymbol{x}_m}&\quad {\sum}_{m=1}^M f_m(\boldsymbol{x}_m) + {\sum}_{l\neq m} \boldsymbol{\lambda}_m^T \boldsymbol{h}_m(\boldsymbol{x}_1,\dots,\boldsymbol{x}_M) \nonumber \\
 \textrm{s.t}& \quad \boldsymbol{h}_m(\boldsymbol{x}_1,\dots,\boldsymbol{x}_M) \leq \boldsymbol{0} \quad \boldsymbol{g}_m(\boldsymbol{x}_m) \leq \boldsymbol{0}
\end{align}
where the first constraint has been separated into different cells compared with that in (\ref{compact form of power subproblem}). The dual variable vector corresponding to the first constraint is denoted by $\boldsymbol{\lambda}_m$.

Fixing the values of all variables and multipliers (indicated by over line) except those in cell m, (\ref{compact form of power subproblem-decomposed}) reduces to
\begin{align} \label{reduced form of power subproblem-decomposed}
 \max_{\boldsymbol{x}_m}&\quad k + f_m(\boldsymbol{x}_m)+{\sum}_{l\neq m}\bar{\boldsymbol{\lambda}}_l^{\rm{T}} \boldsymbol{h}_l({\bf{\bar x}}^m)\nonumber  \\
 \textrm{s.t}& \quad {\bf{\bar x}}^m =[\bar{\boldsymbol{x}}_1,\dots,\bar{\boldsymbol{x}}_{m-1},
 \boldsymbol{x}_m,\bar{\boldsymbol{x}}_{m+1},\dots,\bar{\boldsymbol{x}}_{M}] \nonumber \\
 & \quad \boldsymbol{h}_m({\bf{\bar x}}^m) \leq \boldsymbol{0} \quad \boldsymbol{g}_m(\boldsymbol{x}_m) \leq \boldsymbol{0}
\end{align}
where $k= \sum_{l=1,l\neq m}^M f_l(\bar{\boldsymbol{x}}_l)$ is a constant. Similarly, the reduced problem (\ref{reduced form of power subproblem-decomposed}) can be reduced for every cell $(m=1,\dots,M)$.

%As previously stated, through relaxing all the complicating constraints of others areas and maintaining its own complicating constraints, the optimization problem (\ref{reduced form of power subproblem-decomposed}) is thus reformulated as follows with assumed known optimal values $\boldsymbol{x}_m^*$,$\boldsymbol{\lambda}_m^*$ and $\boldsymbol{\mu}_m^*$:
%\begin{align} \label{reduced form of power subproblem-decomposed-optimal values}
% \max_{\boldsymbol{x}_m}&\quad f_m(\boldsymbol{x}_m) + \sum_{l=1,l\neq m}^M \bar{\boldsymbol{\lambda}}_l^T \boldsymbol{h}_l(\boldsymbol{x}^m) \nonumber  %\\
% \textrm{s.t}&\quad\quad \boldsymbol{h}_m(\boldsymbol{x}^m) \leq \boldsymbol{0} \quad \boldsymbol{g}_m(\boldsymbol{x}^m) \leq \boldsymbol{0}.
% \end{align}global optimization
%where $\boldsymbol{x}^m = (\bar{\boldsymbol{x}}_1^*,\dots,\bar{\boldsymbol{x}}_{m-1}^*,\boldsymbol{x}_m^*,\bar{\boldsymbol{x}}_{m+1}^*,\dots,\bar{\boldsymbol{x}}_{M}^*)$.

To prove that the proposed decomposition method is based on the solutions of these reduced cell subproblems, we revisit the first-order optimality conditions of the problem (\ref{compact form of power subproblem}) which is written as
\begin{align} \label{first-order optimality conditions}
 &\nabla_{\boldsymbol{x}_m} f_m(\boldsymbol{x}_m^*)+ {\sum}_{m=1}^M \nabla_{\boldsymbol{x}_m}^{\rm{T}} \boldsymbol{h}_m(\boldsymbol{x}_1^* ,\dots,\boldsymbol{x}_M^*)\boldsymbol{\lambda}_m^* \nonumber \\
 & +\nabla_{\boldsymbol{x}_m}^{\rm{T}}\boldsymbol{g}_m(\boldsymbol{x}_m^*)\boldsymbol{\mu}_m^* =  \mathbf{0} \nonumber \\
 & \boldsymbol{h}_m(\boldsymbol{x}_1^*,\dots,\boldsymbol{x}_M^*) \leq  \mathbf{0},\quad \quad \boldsymbol{h}_m(\boldsymbol{x}_m^*)^{\rm{T}} \boldsymbol{\lambda}_m^* = 0\nonumber \\
 & \boldsymbol{\lambda}_m^* \geq  \mathbf{0}, \quad \quad \boldsymbol{g}_m(\boldsymbol{x}_m^*) \leq  \mathbf{0},\quad \boldsymbol{g}_m(\boldsymbol{x}_m^*)^{\rm{T}} \boldsymbol{\mu}_m^* = 0\nonumber \\
 & \boldsymbol{\mu}_m^* \geq  \mathbf{0} \quad \quad m=1,\dots,M
\end{align}
These conditions have been constructed using the optimal values $\boldsymbol{x}_m^*$, $\boldsymbol{\lambda}_m^*$ and $\boldsymbol{\mu}_m^*$ which are assumed to be known. The values $\boldsymbol{\lambda}_m^*$ and the values $\boldsymbol{\mu}_m^*$ are the optimal Lagrange multipliers associated with the two constraints in (\ref{compact form of power subproblem}), respectively.
If the first-order optimality conditions of every cell reduced subproblem (\ref{reduced form of power subproblem-decomposed}) $(m=1,\dots,M)$ are stuck together, it can be observed that they are identical to the first-order optimality conditions (\ref{first-order optimality conditions}) of the global problem (\ref{compact form of power subproblem}). Thus, it is obvious that the decomposition mythology holds.

%\noindent \textbf{\underline{Proposed power allocation algorithm:}}

Based on the previously proposed decomposition methodology, removing constants and substituting the objection function and constraints of (\ref{power allocation optimization problem}) into (\ref{reduced form of power subproblem-decomposed}), the original power allocation problem can be decomposed into $M$ subproblems, which are computed in each cell independently. The $m^{\rm{th}}$ subproblem is formulated as
\begin{align} \label{power allocation subproblem}
 \max_{\mathbf{P}_m,R_m}&\quad \omega_m R_m+{\sum}_{l\neq m} \bar{\lambda}_{u,l}(\bar{R}_l-{\sum}_{n:A_{:,:,n}=1}R_{u,l,n}(\mathbf{\bar p}^m)) \nonumber \\
 \textrm{s.t} \ \ & \ \ \ {\sum}_{n:A_{:,:,n}=1}R_{u,m,n}(\mathbf{\bar p}^m) \geq  R_m \ \ \forall{u \in U_m} \nonumber \\
             &\quad {\sum}_{n=1}^N P_{m,n} \leq P_{m,max} \quad\quad \forall{m}.
\end{align} where $\mathbf{p}^m = (\bar{\mathbf{p}}_1,\cdots,\bar{\mathbf{p}}_{m-1},\mathbf{p}_m,\bar{\mathbf{p}}_{m+1},\cdots,\bar{\mathbf{p}}_{M})$, and $\lambda_{u,m}$ is the optimal Lagrange multiplier which guarantees that user rate is larger than minimal value $R_m$ in each cell.

A summary of the proposed decomposition algorithm for power allocation optimization is as follows:
\begin{algorithm}
\caption{Power Allocation Optimization Algorithm}
\SetAlgoNoLine
\SetKwInput{Kw}{Initialize}
\Kw{Initialize the iteration counter $t=0$, and each cell (m = 1, . . . ,M) initializes its variables $\bar{\mathbf{p}}_m,\bar{R}_m$ and parameters $\bar{\lambda}_{u,m}$.}
\Repeat{$\parallel {\rm{vec}}(\mathbf{P}^{t} - \mathbf{P}^{t-1})\parallel < \Psi $ , or $t=T$}
{ \bf{1.} Each cell carries out one iteration for its corresponding subproblem (\ref{power allocation subproblem}), and obtains search directions $\Delta \mathbf{p}_m, \Delta R_m, \Delta \lambda_{u,m}$\;
  \bf{2.} Each cell updates its variables and parameters \\
  $\bar{\mathbf{p}}_m \leftarrow \bar{\mathbf{p}}_m+\Delta \mathbf{p}_m$, $\bar{R}_m \leftarrow \bar{R}_m+\Delta R_m$, $\bar{\lambda}_{u,m} \leftarrow \bar{\lambda}_{u,m}+\Delta \lambda_{u,m}$\;
  \bf{3.} $t=t+1$\;
}
\textbf{Return} $\mathbf{P}^{t}$
\end{algorithm}

The search directions, $\Delta \mathbf{p}_m, \Delta R_m, \Delta \lambda_{u,m}$, for subproblem (\ref{power allocation subproblem}) can be computed independently of each other, allowing a parallel and distributed implementation. This step requires a central agent to coordinate the process, which receives certain information ($\bar{\mathbf{p}}_m,\bar{R}_m, \bar{\lambda}_{u,m}$ after each iteration) from all cells and returns it to the appropriate cells. It can be noted that the information exchanged between the areas and the central agent is little. Unlike other decomposition algorithm, in the proposed algorithm the central agent only distributes information and checks the convergence condition. It does not need to update any information, because this information is updated by the areas of the system, implying a simpler process.

The main difference between the Lagrangian relaxation algorithm and the proposed decomposition one is that Lagrangian relaxation adds all the complicating constraints into the objective function. Therefore it needs auxiliary procedures to update the Lagrange multipliers. On the contrary, the proposed technique does not need any procedure to update the multipliers because this updating is automatic and results directly from the foreign optimization problem. For example, the Lagrangian multipliers $\lambda_l$ $(l\neq m)$ included in the objective of cell $m$ are obtained from cell $l$ 's optimization by keeping its own complicating constraints.
What's more, the proposed approach has the advantage that convergence properties do not require an optimal solution of the subproblems at each iteration of the algorithm. It is enough to perform a single iteration for each subproblem, and then to update variable values. As a consequence, computation times can be significantly reduced with respect to other methods that require the computation of the optimum for the subproblems in order to attain convergence.
\subsection{Subcarrier Allocation Optimization}
When the power allocation is fixed, the remaining optimization problem is to find the optimal subcarrier allocation. It can be directly decomposed into $M$ subproblem, each of which corresponds to problem of maximizing the minimal user rate in cell $m$. The $m^{th}$ is a Mixed Integer Linear Problem (MILP) problem involving both integer and continuous variables, it can be solved by various algorithms such as exhaustive search, implicit enumeration method and branch-and-bound algorithm\cite{Stephen Boyed}.

%Therefore, (\ref{subcarrier allocation optimization problem}) can be solved by the following bisection algorithm.
%\begin{algorithm}
%\caption{Subcarrier Allocation Optimization Algorithm(in $m$th cell)}
%\SetAlgoNoLine
%\SetKwInput{Kw}{Initialize}
%\Kw{Initialize the lower bound $R_m^{\rm{min}}$ and upper bound $R_m^{\rm{max}}$ of minimal user rate in cell $m$}
%\Repeat{$\parallel R_m^{\rm{max}} - R_m^{\rm{min}} \parallel < \Psi_3 $}StephenBoyedStephenBoyedStephenBoyed
%{ \bf{1.} Set $R_m = 0.5(R_m^{\rm{min}}+R_m^{\rm{max}})$\;
%  \bf{2.} Find $\mathbf{A}_m$ satisfying all the constraints in problem (\ref{subcarrier allocation optimization problem})\;
%  \bf{3.} If $\mathbf{A}_m$ exists, $R_m^{\rm{min}} = R_m$, $\mathbf{A}_m^{\rm{opt}}= \mathbf{A}_m$\\
%  \quad  If no $\mathbf{A}_m$ exists, $R_m^{\rm{max}} = R_m$\;}
%\textbf{Return} Optimal subcarrier allocation variables for cell m $\mathbf{A}_m^{\rm{opt}}$
%\end{algorithm}

% ==================================================
\section{Simulation Results}
\label{sec:simulation}

In this section, the performance of the proposed distributed resource allocation algorithm is investigated. The downlink of a cellular OFDMA system is considered with three coordinated cells and 32 subcarriers. The radius of each cell is 40m and there are two users randomly located in each cell. To focus on the performance of proposed algorithm, all weighting coefficients assigned to each cell are assumed to be identical and keep constant. In addition, we set $\Gamma = 0$dB, $\sigma^{2}_{u,m,n}= -60$ dBw and $\Psi_1=\Psi_2=\Psi_3=10^{-1}$. 500 independent channel realizations are carried out to obtain the final results.

\begin{figure}[!ht]
\centering
\includegraphics[width=.3\textwidth]{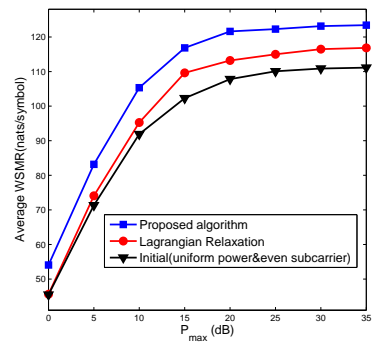}
\caption{Average WSMR of proposed algorithm and the algorithm whose power allocation use Lagrangian relaxation method.}\label{WSMR}
\end{figure}

\begin{figure}[!ht]
\centering
\includegraphics[width=.3\textwidth]{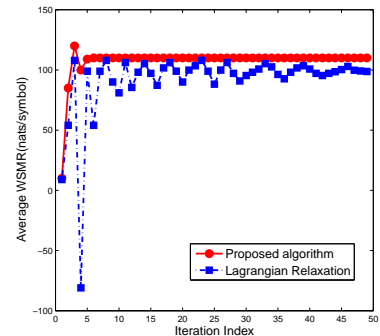}
\caption{Convergence performances of proposed decomposition algorithm and the Lagrangian relaxation method for power allocation.}\label{convergence}
\end{figure}

Fig.\ref{WSMR} compares the average WSMR performance of the proposed algorithm and the one using Lagrangian relaxation method (LR) as a benchmark power allocation algorithm. The average WSMR at initialization which use uniform power allocation and even subcarrier allocation is also shown as a reference. It can be seen that the performance of the proposed distributed resource allocation algorithm is better than that of the algorithm using traditional Lagrangian decomposition method. Fig.\ref{convergence} demonstrates the convergence performance of different algorithm for power allocation. Unlike the slow and oscillating behavior of LR procedure, proposed algorithm converges rapidly and stably. It can be concluded that the proposed decomposition algorithm are more effective than the traditional Lagrangian based algorithms.

% ==================================================
\section{Conclusions}
\label{sec:conclusion}
In this letter, we proposed a distributed algorithm for joint resource allocation in a coordinated multi-cell OFDMA network. The performance criterion termed as WSMR has been introduced to guarantee the fairness among multiple users and cells. Targeting at maximizing WSMR, an iterative algorithm was proposed to solve the joint optimization of power allocation and subcarrier scheduling. At each iteration, the power allocation is updated by applying a modified decomposition methodology which has low complexity and fast convergence speed. On the other hand, the subcarrier allocation is updated by solving a MILP. It is shown that the proposed distributed resource allocation algorithm achieve a better WSMR performance compared compared with the traditional decomposition algorithms. At the same time the proposed power allocation scheme also provides a better convergence performance than the Lagrangian algorithm.

% ==================================================

%\begin{figure}[!ht]
%\centering
%\includegraphics[width=0.3\textwidth]{systemmodel.eps}
%\caption{A distributed cellular OFDMA network.}\label{systemmodel}
%\end{figure}

\end{document}